# THE NOTION OF TIME IN SPECIAL RELATIVITY

Yefim Bakman* and Boris Pogorelsky

*Tel-Aviv University, E-mail: bakman@post.tau.ac.il

**Abstract**. Even though the concepts of time and space are basic concepts of physics, they have not been vouchsafed a constructive definition. As to space, this is no wonder because a correct notion of space cannot be formed in the frame of the existing physical paradigm. We intend to discuss this problem in another article. However, a definition of time can be given on the basis of the simple principle that **each physical magnitude constitutes a feature of some physical carrier** (Costa, 1987). This article deals with identifying that carrier for the feature "time". Finally this article affirms that because of the lack of a definition of time some physical experiments have been erroneously interpreted as signs of time dilation.

## 1. "Time" before special relativity

Before the publication of the theory of special relativity in 1905 time was considered absolutely stable and phrases like "time dilation" were inadmissible. Hence, we can say that time before special relativity was different from time after it.

Let us try and give a definition of time on the basis of the principle "Each physical magnitude is defined by the method it is measured":

> *Definition 1. Time is the magnitude measured by clocks.*

Let us see how this definition works. The period of a pendulum clock is equal to $T = 2\pi\sqrt{l/g}$, where *l* is the pendulum length, *g* is the free fall acceleration. Hence placed in a valley, such a clock will run faster than on the top of a high mountain because the free fall acceleration gets greater when you are going down. Does this mean that "the time in the valley" runs faster? The answer, according to *Definition* 1 must be "yes". But how then do we distinguish between such an acceleration of time from when a regular clock is fast, and what, consequently, does it mean when we say "a clock is fast"? Fast respective to what?

Asking such questions helps us to understand the notion of "time standard" which was in use before the introduction of special relativity theory:

> *Definition 2. The time standard was defined by the angle of the earth's rotation around its axis.*

The fact that people try to synchronize their watches and clocks with the process of the earth's rotation around its axis testifies that this process constitutes the time standard. Consequently, like any standard it is unchangeable by definition and the acceleration of the pendulum clock in the valley does not mean acceleration of time - it only means that the pendulum's swings diverge from the time standard.

## 2. Time as the rate of a process

In the case of length we know that there exists a length standard and we know that **any body** has its particular length. This is quite different in the case of time. Instead of speaking about the time of a process, one must use the term "duration". This is an example how language use can hinder our understanding of a phenomenon. Here language deprives usual processes of the time feature, ascribing them "duration" instead. This results in the illusion that the process of the earth's rotation and the clocks synchronized with it are the only carriers of the feature "time". But in fact **any process** has its time. Compare this with the term "length" which can be used both as the feature of the length standard and as the feature of any other body. Thus, we can offer the following definition of time:

> *Definition 3. Any process has its own time which characterizes the rate of this process.*

The rate of a process can change, so the individual time is changeable.
Consider the earlier mentioned pendulum clock, placed in the valley, and let us try to answer the same question "Does time in the valley run faster?" in the light of the last definition. In the previous section we saw that the clock is fast relative to the time standard. Now, from the viewpoint of *Definition* 3, we know that the phrase "time in the valley" is meaningless because time is not a feature of place, hence **the question is incorrect**. We may say that the time of the pendulum swings runs faster and this is a commonsense phrase like "the length of a body has increased".
Thus, the definitions 1-3 do not contradict one another. On the contrary, together they represent different aspects of the same notion of time:

*Time is the rate of a process. Of all processes, the earth's rotation around its axis was chosen as the time standard. Clocks are attempts to copy the standard but they always measure their own time which is affected by the conditions specific for each clock.*

Now when we have developed the notion of time independent of the language we are ready to consider the time problem in special relativity theory, the problem which resulted from the lack of a constructive time definition.

### 3. The notion of time in special relativity theory

Since special relativity (SR) claims that time dilation is possible, it is clear that SR deals with the particular time of a process, not with the time standard which, by definition, cannot be changed. On the other hand, SR argues that there exists some common time for all bodies which belong to a certain inertial reference frame. This means that each inertial reference frame possesses a time standard of its own which constitutes the carrier of the feature "time" for that specific reference frame. Thus, we conclude that the only explanation that can account for SR's time dilation may be different time standards for different reference frames. But this is not possible either because time dilation is reciprocal (a<b and b<a cannot both be true). Hence, we conclude that the time of SR is not a physical notion - it cannot be a feature of some physical process.

What we described here with regard to time also obtains for the concept length as well. This is, of course, a very bold assertion in the light of the experiments that confirmed the special relativity theory. In the next section we will discuss two of these experiments and we will offer some alternative explanations for their results.

### 4. The long lifetime of quick mesons

The long lifetime of quick mesons compared to slow mesons is one of the few confirmations of SR's conclusions on time dilation in a moving reference frame (Rossi and Hall, [2]). However, we propose another interpretation of the phenomenon.

Let us consider on what the inference of time dilation is based. Time dilation in SR means the following:

    (A) <u>All</u> processes slow down equally (other conditions being equal).

The phrase in parenthesis is important because if, for example, it goes about chemical reaction in the presence of catalytic agents then the deceleration will occur relatively the same reaction with the presence of the same catalytic agents. This means that

(B) Other factors except time can affect the speed of processes.

It follows from Assertion (B) that if an additional factor (except time) affected the speed of the quick mesons decay, the experiment would not prove time dilation. This means that implicitly or explicitly it is assumed that the decay of particles is subjected strictly to one influence, namely the influence of time. Thus, according to SR and contrary to Assertion (B) we have

(C) Time is <u>the unique factor</u> affecting the decay speed.

Indeed, if other factors could affect the decay speed, then it is possible that just they, rather than time, decelerated the quick mesons decay in Rossi and Hall's experiment hence their experiment did not confirm the time delay.

It was Shnoll et al.'s experiment [3-5] that evidenced that Assertion (C) is wrong. For more than three decades these scientists investigated anomalous statistical regularities in a wide range of physical, chemical, and biological processes, from radioactive decay to the rates of biochemical reactions. The evidence points unambiguously to the existence of a previously unknown relationship between fluctuations in the rates of radioactive and other processes in the laboratory, and major astronomical cycles, including the day, month, and year. The implication is, that many phenomena which until now have been regarded as purely statistical - such as the distribution of fluctuations in the momentary rates of radioactivity measured in a sample - are strongly influenced by an astrophysical factor, which varies in time in the same way at all points on the Earth.

Since Shnoll et al. [3-5] discovered the additional factor affecting the decay speed, Rossi and Hall's experiment [2] does not prove time dilation for quick mesons. In GR it is possible to use light speed slowing instead of time dilation [6].

**Conclusion**

The time of SR cannot be a feature of a physical process; therefore it is not a physical concept but rather a handy mathematical symbol.